# Quantum Probes Reduce Measurements: Application to Distributed Grover Algorithm

Iaakov Exman[1] and Efrat Levy[1,2]

[1]Software Engineering Department
Jerusalem College of Engineering
Jerusalem, Israel
and
[2]School of Computer Science and Engineering
Hebrew University of Jerusalem
Jerusalem, Israel
iaakov@jce.ac.il, efratushhh@gmail.com

**Abstract.** Distributed computing seems to be a natural approach to overcome size limitations of quantum computers in terms of number of qubits. But one lacks an efficient distribution approach to deal systematically with potential algorithms. This work proposes suitable quantum probes to reduce measurements in distributed sub-systems. Probes are introduced in the context of distributed Grover search. These are shown to be indeed efficient. The latter can be generalized to be applicable to other distributed algorithms of interest.

**Keywords:** quantum probes, distributed algorithms, measurement reduction, Grover search.

## 1 Introduction

The well-known Grover search algorithm [3] is one of the prominent algorithms in quantum computing. It allows quantum search of an unstructured database in time $O(\sqrt{N})$, which is significantly faster than the equivalent classical result $O(N)$.

Essentially, after an initialization stage, Grover search, given an integer input *x,* works by iterative application of:

- an oracle $O$ that recognizes a solution – a function *f(x)* which equals 1 if *x* is a solution of the search problem, and equals 0 otherwise:

$$|x> \xrightarrow{O} (-1)^{f(x)} |x> \qquad (1)$$

- an operator that performs an inversion about the average.

Finally the N qubits are measured.

Grover search can be distributed into smaller sub-systems that enable execution in parallel, while fitting to the quantum computer size limitations of the current implementation technology. Grover search distribution approaches may be inefficient, consuming resources that waste the advantages of parallelism.

This work proposes quantum probes, to be composed with sub-systems, which contribute to distribution efficiency and reduce the number of measurements necessary to attain the algorithm results.

Next we further motivate the need for distributed algorithms and overview the relevant literature.

## 1.1 Motivation for Distributed Algorithms

The current state of the art in quantum computers – irrespective of the implementation technology – is still very limited in size, in terms of numbers of qubits.

An actual implementation of Grover search has been done by Vandersypen et al. [15] in a 3 qubits computer consisting of $^{13}C$ labeled $CHFBr_2$ molecules, in which the three weakly coupled spin-1/2 nuclei behave as the bits and are manipulated using magnetic resonance techniques.

The largest implemented quantum computer nowadays – cf. ref. [10] by Knill et al. – features 7 qubits, provided by the nuclei of crotonic acid. This implementation also makes use of nuclear magnetic resonance (NMR) techniques.

There have been estimates of increasing of quantum computer sizes. Taking an optimistic point of view, there are references up to a few hundred qubits. Even assuming that numbers of qubits will significantly increase in the coming future, it is reasonable to assume that one will always face size limitations.

Distributed algorithms should enable the combination of existing quantum computers – whatever their size – to solve problems of larger sizes with potential practical interest.

## 1.2 Related Work

Quantum search was proposed in Grover's original paper of 1996 [3]. Grover continued to investigate quantum search in subsequent papers, such as search of structured problems [5] – an issue also investigated by Hogg [7] – and search of arbitrarily large databases by a single query [6]. Grover's search algorithm bounds [1] and optimality [17] have been established.

Variations on quantum search have been proposed for mixed states by Kenigsberg [9], for search without entanglement, by Lloyd [12] and by Meyer [13], and for an arbitrary set of initial states by Shimoni, Shapira and Biham [14].

Quantum computers with a variety of technologies have been proposed. Most of the actually implemented quantum computers have very few qubits. Proposals have also been made on how to extend the number of qubits, e.g. Krojanski and Suter [11] by wide NMR quantum registers, and Brickman et al. in scalable systems [2].

Distributed quantum computing has been focused from various points of view. Yimsiriwattana and Lomonaco [16] proposed cat-entangler and cat-disentangler as primitive operations which can be used to implement non-local operations, and mainly to distribute a control qubit among computers.

Ancilla qubits have been called probe qubits by Zeng et al. [19]. Although displaying similarities to our quantum probes, they have been used in a different context, viz. to measure parity of N-qubit states. This same application was also the subject of the more recent work of Ionicioiu et al. [8].

Zalka [18] also questioned the practicality of the issue of searching an actual database, with an inconclusive outcome.

In the remaining sections of the paper, we describe distributed approaches to Grover Search (in section 2), precisely formulate the notion of quantum probes and its application to Grover distribution (in section 3), and conclude with a discussion (in section 4).

## 2 Distribution Approaches to Grover Algorithm

There are a few possible approaches to distribute the Grover algorithm. In order to classify them, we first characterize in a general way, the common stages of the diverse approaches.

## 2.1 Stages of Distributed Grover Algorithm

To distribute an algorithm one needs, instead of a single system executing an original non-distributed algorithm on an input with *N* items, to have M distributed sub-systems, each working on a suitable sub-set of the whole input. In general, the sizes of different sub-systems may be different.

One can describe a general distributed algorithm by three stages:

1. *Distribution* – in this stage each sub-system is initialized with its respective sub-set of the input;
2. *Operation in sub-system* – the original non-distributed algorithm is performed in each sub-system;
3. *Merging and Decision* – the sub-system results are collected and a merging and decision procedure is performed to obtain the final result.

For the case of a distributed Grover algorithm the input is a database with $N = 2^k$ items. One must be careful in the *Distribution* stage to take into account the offset between the index in the whole database and the index of each sub-system.

In the *Operation* stage the operation itself is the regular Grover search, with an oracle implementation suitable for the sub-system size. Thus, each sub-system after measurement outputs a result, whether the intermediate result is a true final result or not.

For the *Merging and Decision* stage we shall temporarily assume that one has a single valid result for the whole database, meaning that the final result contains just one and only one item. This assumption can be easily relaxed to treat result multiplicity.

Diverse distributing approaches differ mainly by the transition between stage 2 and stage 3 and by the nature of the merging and decision procedure, as seen next.

## 2.2 Semi-classical Approaches

We call a distribution approach semi-classical, when in stage 2, at the end of the Grover search run in the sub-system, its intermediate result is measured.

After the measurement occurs in stage 2, one has various alternative ways to decide which intermediate sub-system result is a correct one. Here are some examples:

- Each of the intermediate results is verified classically by an oracle application in stage 3;
- The Grover search runs in each sub-system are repeated a few times – less than $\sim\sqrt{N}$ by considerations of the birthday paradox. If the outcomes of the consecutive search runs are repetitive, this is a true final result.

It is a desirable aim to minimize measurements and continue as far as possible within a quantum approach. To this end we discuss quantum probes in the next section.

## 3 Quantum Probes

A quantum probe is an ancillary system containing a small number of qubits, which is composed with a larger system, in order to obtain a computation result by means of a reduced number of measurements. The number of measurements is reduced relative to an equivalent result obtained from the larger system without the ancilla.

### 3.1 Detailed Characterization of Quantum Probes

A typical quantum probe is used in the following way:

1. *Composition* – a larger system $|\Psi\rangle$ is composed with the quantum probe $|q_p\rangle$:

$$|\Psi, q_p\rangle = |\Psi\rangle \otimes |q_p\rangle \qquad (2)$$

2. *Unitary operation* – a unitary operation is applied to the composed system:

$$|\Psi, q_p\rangle \xrightarrow{U_f} |\Psi, f(x) \oplus q_p\rangle \qquad (3)$$

3. *Measurement* – the quantum probe is measured, instead of measuring the original larger system.

## 3.2 Application to Distributed Grover

The purpose of the quantum probe in the distributed Grover algorithm is to reduce the number of measurements in each sub-system.

At the end of the Grover algorithm, the state of a sub-system that does not have a search solution is $\sum_i |i\rangle$, where $i$ stands for any item index. On the other hand, the state of a sub-system having a true solution $i_0$ is given by:

$$|\Psi\rangle = a|i_0\rangle + b \sum_{i \neq i_0} |i\rangle \qquad (4)$$

where the coefficients $a$ and $b$ are such that $a \approx 1$ and $b \approx 0$.

Let us choose a quantum probe with a single qubit in the state $|0\rangle$. After *Composition* of $|\Psi\rangle$ with the quantum probe and application of the oracle function, one obtains:

$$|\Psi, q_p\rangle = a|i_0, 1\rangle + b \sum_{i \neq i_0} |i, 0\rangle \qquad (5)$$

*Measurement* of the quantum probe alone gives the $|1\rangle$ state with high probability only for the sub-system containing the search solution. Thus, we measure only one qubit per sub-system instead of all qubits in each sub-system.

## 3.3 Distributed Grover Efficiency

Straightforward calculation of the overall number of operations applied in the distributed Grover search with the above quantum probes shows that it is indeed efficient. It consumes less space – other verifiers such as found in semi-classical approaches are not needed in such a stage – and also less time as one measures O(*1*) qubits in each sub-system.

In particular:

- *Grover search in sub-systems* – since all sub-systems run in parallel, including the final measurement of the quantum probe, the overall run-time is $O(\sqrt{v})$, where $v$ is now the size of each sub-system;
- *Finding the correct sub-system* – this amounts to finding the position of the only classical bit with value 1 in a binary number with $M$ bits, where $M = \left(\frac{N}{v}\right)$ is the number of sub-systems, taking O(log(*M*));

To the above computations one should add the initial database sub-sets distribution to sub-systems and the final recovery of the correct result from the correct sub-system. Both of these involve calculating the offset between the whole database index to the respective sub-system index, only once at the initialization and once at the termination of the whole computation. It is reasonable to assume that these calculations take a negligible constant time, compared to the Grover search in sub-systems.

One concludes that due to parallel execution of sub-systems and to the use of quantum probes, the overall distributed Grover search is more run-time efficient than the original sequential Grover search.

## 4 Discussion

This work has considered several approaches to distribution of Grover search in smaller sub-systems. Semi-classical approaches besides having larger numbers of measurements, may involve repeated rounds of Grover iterations, which is very time consuming.

Our proposed approach uses quantum probes – a few ancilla qubits to be composed with Grover sub-systems. Quantum probes, besides involving a single additional oracle application, are efficient in terms of reduced numbers of measurements.

A possible disadvantage of quantum distributed approaches in general, relative to the sequential quantum counterparts, refers to the storage of the information needed to represent an N-qubit system. In a classical computer this would require the storage of $2^n$ complex coefficients, while qubits in a quantum computer can hold exponentially more information. Distribution into smaller quantum sub-systems wastes this advantage to a certain extent, which depends on the quantum sub-system size.

### 4.1 Open Issues

An important issue is the generalization of the distributed Grover with the quantum probe technique to other algorithms.

The use of the quantum probe implies that for the distributed Grover search we can get the true sub-system without changing its *original* state. A first possible generalization is that every *query* quantum distributed algorithm should give an answer without a measurement of the original state, i.e. the state before the composition with the quantum probe.

### 4.2 Main Contribution

The main contribution of this paper is the application of quantum probes – a few qubits ancilla composed with a system, reducing measurements in a computation – to obtain results of quantum distributed computing in a time-efficient and elegant way.

## References

[1] Boyer, M., Brassard, G., Hoyer, P. and Tapp, A., "Tight bounds on quantum searching", Fortsch. Phys. Vol. 46 pp. 493-506 (1998), also Proc. PhysComp (1996), and quant-ph/9605034
[2] Brickman, K.A., Haljan, P.C., Lee, P.J., Acton, M., Delauriers, L. and Monroe, C., "Implementation of Grover's quantum search algorithm in scalable system", FOCUS Center and Department of Physics, University of Michigan, USA pp. 2-4 (2005).
[3] Grover, L. K., "A fast quantum mechanical algorithm for database search", Proc. 28th STOC Annual Symposium on the Theory of Computing, pp. 212-219, (1996).
[4] Grover, L. K., "Quantum Mechanics helps in searching for a needle in a haystack", Phys. Rev. Letters, vol. 78, pp. 325-328, (1997), also quant-ph/9605043
[5] Grover, L.K., "Quantum search on structured problems", in 1st NASA Conf. on quantum computation and quantum communication, Palm Springs, CA, (February 1998), also quant-ph/9802035.
[6] Grover, L. K., "Quantum computers can search arbitrarily large databases by a single query", Phys. Rev. Letters, vol. 79, pp. 4709-4712 (1997).
[7] Hogg, T., "A framework for structured quantum search", quant-ph/9605043.
[8] Ionicioiu, R., Popescu, A.E., Munro, W.J. and Spiller, T.P., "Generalized Parity Measurements", Phys. Rev. A 78, 052326 (2008), also arXiv:quant-ph/0806.0982 (2008).
[9] Kenigsberg, D., "Grover's Quantum Search Algorithm and Mixed States", M.Sc. Thesis, Computer Science Department, Technion – Israel Institute of Technology (2001).
[10] Knill, E., Laflamme, R., Martinez, R. and Tseng, C.-H., "A Cat-State Benchmark on a Seven Bit Quantum Computer", arXiv:quant-ph/9908051 (1999).
[11] Krojanski, H.G. and Suter, D., "Scaling of Decoherence in Wide NMR Quantum Registers", *Phys. Rev. Letters*, Vol. 93, pp. 090501 (2004).
[12] Lloyd, S., "Quantum search without entanglement", *Phys. Rev.*, A61, pp. 1-7 (1999).
[13] Meyer, D. A., "Sophisticated quantum search without entanglement", *Phys. Rev. Letters*, Vol. 85, pp. 1-6 (2000).
[14] Shimoni, Y., Shapira, D. and Biham, O., "Characterization of pure quantum states of multiple qubits using the Groverian entanglement measure", Phys. Rev. A 69, 062303 (2004), also arXiv:quant-ph/0309.0362 (2003).
[15] Vandersypen, L.M. K., Steffen, M., Breyta, G., Yannoni, C. S., Sherwood, M. H. and Chuang, I. L., "Implementation of a three-quantum-bit search algorithm", Nature (London) Vol. 414, 883 (2001), also arXiv:quant-ph/9910075 (2000).
[16] Yimsiriwattana, A. and Lomonaco Jr., S.J., "Generalized GHZ States and Distributed Quantum Computing", Arxiv quant-ph/0403146, 2004
[17] Zalka, C., "Grover's quantum searching algorithm is optimal", Phys. Rev. A, vol. 60, pp. 2746-2751, (1999), also quant-ph/9711070.
[18] Zalka, C., "Could Grover's quantum algorithm help in searching an actual database?", arXiv:quant-ph/9901068 (rev. Dec 2005).
[19] Zeng, B., Zhou, D. L. and You L. "Measuring the Parity of an *N*-Qubit State", Phys. Rev. Letters, Vol. 95, 110502 (2005).